\documentclass[aps,prl,superscriptaddress,showpacs,floatfix,twocolumn]{revtex4}

\usepackage{graphicx}   
\usepackage{dcolumn}
\usepackage{bm}
\usepackage{amssymb}

\def\pT{\mbox{$p_T$}}

\begin{document} 

\title{Azimuthal anisotropy of $\pi^0$ production in Au+Au collisions at
$\sqrt{s_{\rm NN}}=200$~GeV:  Path-length dependence of
jet-quenching and the role of initial geometry}

\newcommand{\abilene}{Abilene Christian University, Abilene, Texas 79699, USA}
\newcommand{\banaras}{Department of Physics, Banaras Hindu University, Varanasi 221005, India}
\newcommand{\barc}{Bhabha Atomic Research Centre, Bombay 400 085, India}
\newcommand{\bnlcoll}{Collider-Accelerator Department, Brookhaven National Laboratory, Upton, New York 11973-5000, USA}
\newcommand{\bnlphys}{Physics Department, Brookhaven National Laboratory, Upton, New York 11973-5000, USA}
\newcommand{\caucr}{University of California - Riverside, Riverside, California 92521, USA}
\newcommand{\charlesczech}{Charles University, Ovocn\'{y} trh 5, Praha 1, 116 36, Prague, Czech Republic}
\newcommand{\chonbuk}{Chonbuk National University, Jeonju 561-756, Korea}
\newcommand{\ciae}{China Institute of Atomic Energy (CIAE), Beijing, People's Republic of China}
\newcommand{\cns}{Center for Nuclear Study, Graduate School of Science, University of Tokyo, 7-3-1 Hongo, Bunkyo, Tokyo 113-0033, Japan}
\newcommand{\colorado}{University of Colorado, Boulder, Colorado 80309, USA}
\newcommand{\columbia}{Columbia University, New York, New York 10027 and Nevis Laboratories, Irvington, New York 10533, USA}
\newcommand{\czechtech}{Czech Technical University, Zikova 4, 166 36 Prague 6, Czech Republic}
\newcommand{\dapnia}{Dapnia, CEA Saclay, F-91191, Gif-sur-Yvette, France}
\newcommand{\debrecen}{Debrecen University, H-4010 Debrecen, Egyetem t{\'e}r 1, Hungary}
\newcommand{\elte}{ELTE, E{\"o}tv{\"o}s Lor{\'a}nd University, H - 1117 Budapest, P{\'a}zm{\'a}ny P. s. 1/A, Hungary}
\newcommand{\ewha}{Ewha Womans University, Seoul 120-750, Korea}
\newcommand{\fit}{Florida Institute of Technology, Melbourne, Florida 32901, USA}
\newcommand{\fsu}{Florida State University, Tallahassee, Florida 32306, USA}
\newcommand{\gsu}{Georgia State University, Atlanta, Georgia 30303, USA}
\newcommand{\hiroshima}{Hiroshima University, Kagamiyama, Higashi-Hiroshima 739-8526, Japan}
\newcommand{\ihepprot}{IHEP Protvino, State Research Center of Russian Federation, Institute for High Energy Physics, Protvino, 142281, Russia}
\newcommand{\illuiuc}{University of Illinois at Urbana-Champaign, Urbana, Illinois 61801, USA}
\newcommand{\instpasczech}{Institute of Physics, Academy of Sciences of the Czech Republic, Na Slovance 2, 182 21 Prague 8, Czech Republic}
\newcommand{\isu}{Iowa State University, Ames, Iowa 50011, USA}
\newcommand{\jinrdubna}{Joint Institute for Nuclear Research, 141980 Dubna, Moscow Region, Russia}
\newcommand{\jyvaskyla}{Helsinki Institute of Physics and University of Jyv{\"a}skyl{\"a}, P.O.Box 35, FI-40014 Jyv{\"a}skyl{\"a}, Finland}
\newcommand{\kek}{KEK, High Energy Accelerator Research Organization, Tsukuba, Ibaraki 305-0801, Japan}
\newcommand{\kfki}{KFKI Research Institute for Particle and Nuclear Physics of the Hungarian Academy of Sciences (MTA KFKI RMKI), H-1525 Budapest 114, POBox 49, Budapest, Hungary}
\newcommand{\korea}{Korea University, Seoul 136-701, Korea}
\newcommand{\kurchatov}{Russian Research Center ``Kurchatov Institute", Moscow, Russia}
\newcommand{\kyoto}{Kyoto University, Kyoto 606-8502, Japan}
\newcommand{\labllr}{Laboratoire Leprince-Ringuet, Ecole Polytechnique, CNRS-IN2P3, Route de Saclay, F-91128, Palaiseau, France}
\newcommand{\lawllnl}{Lawrence Livermore National Laboratory, Livermore, California 94550, USA}
\newcommand{\losalamos}{Los Alamos National Laboratory, Los Alamos, New Mexico 87545, USA}
\newcommand{\lpc}{LPC, Universit{\'e} Blaise Pascal, CNRS-IN2P3, Clermont-Fd, 63177 Aubiere Cedex, France}
\newcommand{\lund}{Department of Physics, Lund University, Box 118, SE-221 00 Lund, Sweden}
\newcommand{\maryland}{University of Maryland, College Park, Maryland 20742, USA}
\newcommand{\mass}{Department of Physics, University of Massachusetts, Amherst, Massachusetts 01003-9337, USA }
\newcommand{\muenster}{Institut fur Kernphysik, University of Muenster, D-48149 Muenster, Germany}
\newcommand{\muhlenberg}{Muhlenberg College, Allentown, Pennsylvania 18104-5586, USA}
\newcommand{\myongji}{Myongji University, Yongin, Kyonggido 449-728, Korea}
\newcommand{\nagasaki}{Nagasaki Institute of Applied Science, Nagasaki-shi, Nagasaki 851-0193, Japan}
\newcommand{\newmex}{University of New Mexico, Albuquerque, New Mexico 87131, USA }
\newcommand{\nmsu}{New Mexico State University, Las Cruces, New Mexico 88003, USA}
\newcommand{\ornl}{Oak Ridge National Laboratory, Oak Ridge, Tennessee 37831, USA}
\newcommand{\orsay}{IPN-Orsay, Universite Paris Sud, CNRS-IN2P3, BP1, F-91406, Orsay, France}
\newcommand{\peking}{Peking University, Beijing, People's Republic of China}
\newcommand{\pnpi}{PNPI, Petersburg Nuclear Physics Institute, Gatchina, Leningrad region, 188300, Russia}
\newcommand{\riken}{RIKEN Nishina Center for Accelerator-Based Science, Wako, Saitama 351-0198, JAPAN}
\newcommand{\rikjrbrc}{RIKEN BNL Research Center, Brookhaven National Laboratory, Upton, New York 11973-5000, USA}
\newcommand{\rikkyo}{Physics Department, Rikkyo University, 3-34-1 Nishi-Ikebukuro, Toshima, Tokyo 171-8501, Japan}
\newcommand{\saispbstu}{Saint Petersburg State Polytechnic University, St. Petersburg, Russia}
\newcommand{\saopaulo}{Universidade de S{\~a}o Paulo, Instituto de F\'{\i}sica, Caixa Postal 66318, S{\~a}o Paulo CEP05315-970, Brazil}
\newcommand{\seoulnat}{Seoul National University, Seoul 151-742, Korea}
\newcommand{\stonybrkc}{Chemistry Department, Stony Brook University, Stony Brook, SUNY, New York 11794-3400, USA}
\newcommand{\stonycrkp}{Department of Physics and Astronomy, Stony Brook University, SUNY, Stony Brook, New York 11794, USA}
\newcommand{\tenn}{University of Tennessee, Knoxville, Tennessee 37996, USA}
\newcommand{\titech}{Department of Physics, Tokyo Institute of Technology, Oh-okayama, Meguro, Tokyo 152-8551, Japan}
\newcommand{\tsukuba}{Institute of Physics, University of Tsukuba, Tsukuba, Ibaraki 305, Japan}
\newcommand{\vandy}{Vanderbilt University, Nashville, Tennessee 37235, USA}
\newcommand{\waseda}{Waseda University, Advanced Research Institute for Science and Engineering, 17 Kikui-cho, Shinjuku-ku, Tokyo 162-0044, Japan}
\newcommand{\weizmann}{Weizmann Institute, Rehovot 76100, Israel}
\newcommand{\yonsei}{Yonsei University, IPAP, Seoul 120-749, Korea}
\affiliation{\abilene}
\affiliation{\banaras}
\affiliation{\barc}
\affiliation{\bnlcoll}
\affiliation{\bnlphys}
\affiliation{\caucr}
\affiliation{\charlesczech}
\affiliation{\chonbuk}
\affiliation{\ciae}
\affiliation{\cns}
\affiliation{\colorado}
\affiliation{\columbia}
\affiliation{\czechtech}
\affiliation{\dapnia}
\affiliation{\debrecen}
\affiliation{\elte}
\affiliation{\ewha}
\affiliation{\fit}
\affiliation{\fsu}
\affiliation{\gsu}
\affiliation{\hiroshima}
\affiliation{\ihepprot}
\affiliation{\illuiuc}
\affiliation{\instpasczech}
\affiliation{\isu}
\affiliation{\jinrdubna}
\affiliation{\jyvaskyla}
\affiliation{\kek}
\affiliation{\kfki}
\affiliation{\korea}
\affiliation{\kurchatov}
\affiliation{\kyoto}
\affiliation{\labllr}
\affiliation{\lawllnl}
\affiliation{\losalamos}
\affiliation{\lpc}
\affiliation{\lund}
\affiliation{\maryland}
\affiliation{\mass}
\affiliation{\muenster}
\affiliation{\muhlenberg}
\affiliation{\myongji}
\affiliation{\nagasaki}
\affiliation{\newmex}
\affiliation{\nmsu}
\affiliation{\ornl}
\affiliation{\orsay}
\affiliation{\peking}
\affiliation{\pnpi}
\affiliation{\riken}
\affiliation{\rikjrbrc}
\affiliation{\rikkyo}
\affiliation{\saispbstu}
\affiliation{\saopaulo}
\affiliation{\seoulnat}
\affiliation{\stonybrkc}
\affiliation{\stonycrkp}
\affiliation{\tenn}
\affiliation{\titech}
\affiliation{\tsukuba}
\affiliation{\vandy}
\affiliation{\waseda}
\affiliation{\weizmann}
\affiliation{\yonsei}
\author{A.~Adare} \affiliation{\colorado}
\author{S.~Afanasiev} \affiliation{\jinrdubna}
\author{C.~Aidala} \affiliation{\mass}
\author{N.N.~Ajitanand} \affiliation{\stonybrkc}
\author{Y.~Akiba} \affiliation{\riken} \affiliation{\rikjrbrc}
\author{H.~Al-Bataineh} \affiliation{\nmsu}
\author{J.~Alexander} \affiliation{\stonybrkc}
\author{K.~Aoki} \affiliation{\kyoto} \affiliation{\riken}
\author{Y.~Aramaki} \affiliation{\cns}
\author{E.T.~Atomssa} \affiliation{\labllr}
% ...
\collaboration{PHENIX Collaboration} \noaffiliation

\date{\today}

\begin{abstract}


We have measured the azimuthal anisotropy of $\pi^0$ production 
for $1<\pT<18$ GeV/$c$ for Au$+$Au collisions at $\sqrt{s_{\rm
NN}}=200$~GeV. The observed anisotropy shows a gradual decrease for
$3 \alt \pT \alt 7-10$ GeV/$c$, but remains positive beyond 10
GeV/$c$. The magnitude of this anisotropy is under-predicted, up to
at least $\sim10$ GeV/$c$, by current perturbative QCD (pQCD)
energy-loss model calculations.  An estimate of the increase in
anisotropy expected from initial-geometry modification due to gluon
saturation effects and fluctuations is insufficient to account for
this discrepancy.  Calculations that implement a path length
dependence steeper than what is implied by current pQCD energy-loss
models show reasonable agreement with the data.

\end{abstract}


\maketitle 


A central goal of high-energy nuclear physics is to understand the
properties of the strongly-coupled Quark Gluon Plasma (sQGP), a new
form of nuclear matter identified at the Relativistic Heavy Ion
Collider (RHIC)~\cite{RHIC}.  An important tool for this goal is jet
quenching or the suppression of high transverse momentum ($\pT$)
hadron yields as a result of in-medium energy loss of high $\pT$
jets~\cite{Gyulassy:2003mc}.  Such suppression was first observed in
measurements of the nuclear modification factor for single hadron
yields $R_{\rm AA}= \frac{dN_{\rm AA}}{\langle T_{\rm AA}\rangle
d\sigma_{\rm pp}}$, where $dN_{\rm AA}$ is the differential yield
in Au+Au collisions, $d\sigma_{\rm pp}$ is the differential cross
section in p+p collisions for given $\pT$, and $\langle T_{\rm
AA}\rangle$ is the nuclear overlap integral for given Au+Au
centrality bin~\cite{Adcox:2001jp}.  Later on this effect was also
observed in measurements of dihadron~\cite{Adler:2002tq} and
$\gamma$-hadron correlations~\cite{Adare:2009vd}.

Current theoretical descriptions of jet quenching are commonly
based on a pertubative QCD (pQCD) framework~\cite{Majumder:2010qh},
which assumes that the coupling of jets with the medium is weak,
even though the medium itself is strongly-coupled.  Prompted by the
large amount of experimental data from RHIC, several sophisticated
pQCD-based models have been developed in the last
decade~\cite{Gyulassy:2003mc,Majumder:2010qh}.  These models have
provided initial estimates of the properties of the sQGP, such as
the momentum broadening per mean free path, $\hat{q}=\langle
k_T^2\rangle/\lambda$, and the energy loss per unit length,
$dE/dl$~\cite{Adare:2008cg,Bass:2008rv,Majumder:2010qh}.

\begin{figure*}[t]
\centering
\includegraphics[width=1\linewidth]{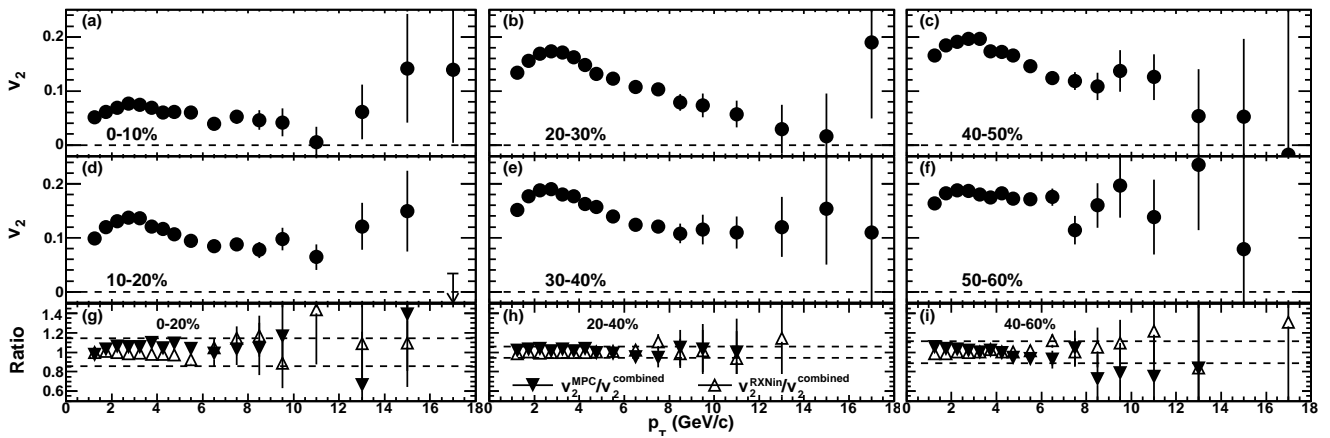}	
\caption{ \label{fig:1} (a)-(f): $\pi^0$ $v_2$ using reaction plane
determined with MPC and RXN$_{\rm in}$ combined as a function of
$\pT$ for different centralities.  (g)-(i): ratios of $v_2$ measured
separately using MPC (solid triangles) and RXN$_{\rm in}$ (open
triangles) to combined result; the dashed lines indicate the
systematic error.}
\end{figure*}

Despite these successes, the pQCD description of jet quenching
faces several challenges (see Ref~\cite{Muller:2008zzm}).  Besides a
large discrepancy among models of extracted medium properties such
as $\hat{q}$~\cite{Bass:2008rv}, the energy-loss models also
disagree in their predictions of the azimuthal anisotropy of high
$\pT$ hadrons~\cite{Bass:2008rv}.  The latter characterizes hadron
emission relative to the reaction plane angle ($\Psi_{RP}$),
$dN/d(\phi-\Psi_{\rm RP}) \propto (1+2v_2\cos(2(\phi-\Psi_{\rm
RP})))$.  Such azimuthal anisotropy ensues because the hadron yield
is more suppressed along the long axis of the almond-shaped
fireball than the short axis.  Thus the magnitude of the anisotropy,
$v_2$, is sensitive to the path length ($l$) dependence of energy
loss, which scales as $\Delta E \sim l$ for collisional energy
loss~\cite{Peigne:2008wu}, $\Delta E \sim l^2$ for coherent
radiative energy loss~\cite{Peigne:2008wu}, and $\Delta E\sim l^3$
for a non-perturbative energy loss calculation using AdS/CFT
gravity-gauge dual theory~\cite{Gubser:2008as}.  However, our
ability to probe such $l$ dependences hinges not only on precision
data at high $\pT$, but also on a good understanding of the role of
the initial collision geometry.  One geometry commonly used in
energy-loss models is based on the Optical Glauber
model~\cite{Miller:2007ri}, which assumes a smooth Woods-Saxon
nuclear geometry.  Such geometry ignores the event-by-event shape
distortion due to spacial fluctuations of participating
nucleons~\cite{Alver:2008zza}, and a possible overall shape
distortion due to gluon saturation effects, {\it i.e.}~the so
called CGC geometry~\cite{Drescher:2006pi}.  These effects have been
shown to be important (up to 15-30\% each) for elliptic flow at low
$\pT$~\cite{Luzum:2008cw,Hirano:2009ah}.  However, their influences
on jet quenching $v_2$ at high $\pT$ are not well studied to date.


In this Letter we present a new measurement of the $\pi^0$
anisotropy in $\sqrt{s_{\rm NN}} = 200$~GeV Au+Au collisions.  This
measurement complements our prior results~\cite{Adler:2005rg,
Adler:2006bw, Afanasiev:2009iv}, but significantly increases both
the $\pT$ reach and the statistical precision above 6 GeV/$c$,
allowing for quantitative comparisons to energy-loss models, as
well as detailed investigations of the role of the initial
collision geometry.

Results were obtained from $\sim3.5\times 10^9$ minimum bias events
taken in 2007.  Event centrality was determined by the number of
charged particles detected in the Beam-Beam Counters (BBC,
$3.0<|\eta|<3.9$).  A Monte-Carlo (MC) Glauber
model~\cite{Miller:2007ri} was used to estimate the average number
of participating nucleons ($N_{\rm part}$) and $\langle T_{\rm
AA}\rangle$ for each centrality class.

Previous PHENIX analyses~\cite{Afanasiev:2009iv} estimated the RP
using the charged particles detected in the BBC.  Several new
detectors, installed symmetrically on both sides of the beam line,
provided additional RP measurements in 2007: the Muon Piston
Calorimeters (MPC, $3.1<|\eta|<3.9$) and the Reaction Plane
detectors in two $\eta$ ranges, RXN$_{\rm in}$ (RXN$_{\rm out}$) in
$1.5 (1.0) < |\eta| < 2.8 (1.5)$.  Each MPC is equipped with
PbWO$_4$ crystal scintillators to detect both charged and neutral
particles.  Each RXN consists of 12 azimuthally segmented paddle
scintillators.  This analysis estimates the RP angle using both the
MPC and RXN$_{\rm in}$ to provide good resolution, while minimizing
the potential biases from jets and dijets~\cite{Adare:2008cqb}.  The
error on the RP angle $\Delta\Psi$, and the RP dispersion factor
$\sigma_{\rm RP}=\langle\cos2\Delta\Psi\rangle$ are estimated by
the sub-event method~\cite{Afanasiev:2009iv}, giving $\sigma_{\rm
RP}\sim 0.52$ and 0.73 in the central and mid-central collisions,
respectively, which is $\sim80$\% better than that for the BBCs.
The large dataset and improved $\sigma_{\rm RP}$ give an equivalent
of $\sim15$ fold increase in statistics over the previous
measurement of $v_2$~\cite{Afanasiev:2009iv}.

The methodology for $v_2$ extraction follows our previous
work~\cite{Afanasiev:2009iv}.  We reconstruct the neutral pions via
the $\pi^{0}\to\gamma+\gamma$ decay channel with photons detected
in the Electromagnetic Calorimeter (EMC, $|\eta|<0.35$).  We apply
shower shape and pair asymmetry cuts to reduce the combinatorial
background.  The remaining background is subtracted by the mixed
event method~\cite{Afanasiev:2009iv}.  The azimuthal distribution of
the $\pi^{0}$ yields relative to the estimated RP angle,
$\Delta\phi=\phi-\Psi_{\rm RP}$, is divided into 6 bins in the
interval of [$0,\pi/2$], and fit to $N_{0}(1+2v_{2}^{\rm
raw}\cos(2\Delta\phi))$.  The $v_2$ is then obtained by applying the
dispersion correction $v_{2}=v_{2}^{\rm raw}/\sigma_{\rm RP}$ for
each centrality and $p_{T}$ selection.  The main sources of
systematic uncertainties come from $\sigma_{\rm RP}$ and $v_2^{\rm
raw}$.  The former is estimated by comparing measurements from
different RP detectors, giving $\sim10\%$ for central and
peripheral collisions and $\sim5\%$ for mid-central collisions.  The
latter accounts for the dependence of $v_2$ on $\pi^0$
identification cuts, as well as the variation among different
sectors of EMC and different run groups, and is correlated in
$\pT$; it is estimated to be 10\% for central collisions and $3\%$
for other collisions.

Figure~\ref{fig:1} (a)-(f) shows $v_2(\pT)$ for six centrality
bins, spanning 1-18 GeV/$c$.  In the 10-50\% centrality range, where
the signal is large and the uncertainty is small, the $v_2$ values
above 3 GeV/$c$ indicate a slow decrease up to 7-10 GeV/$c$, and
remain significantly above zero at higher $\pT$.  The ratios in
Fig.~\ref{fig:1} (g)-(i) confirm the consistency of $v_2$ measured
using the RP from the MPC or the RXN$_{\rm in}$ and imply that
the influence of rapidity dependent jet bias to the RP, if any, is
within the statistical or systematic uncertainty of the
measurement.

\begin{figure}[ht]
\includegraphics[width=1.0\linewidth]{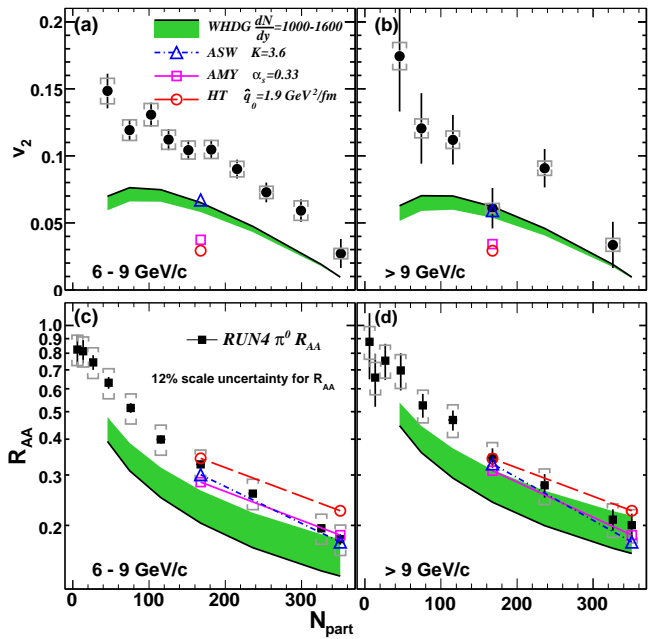}
\caption{\label{fig:2} (Color online) (a)-(b) $v_2$ vs $N_{\rm part}$ in two $\pT$ ranges; (c)-(d) $R_{\rm AA}$ vs $N_{\rm part}$ in same $\pT$ ranges.
Each are compared with four pQCD models from~\cite{Bass:2008rv} (AMY, HT, ASW) and ~\cite{Wicks:2005gt}
(WHDG).  Log-scale is used for $R_{\rm AA}$ to better visualize various model calculations.
Note that the $\frac{dN}{dg}$=1000 of WHDG corresponds to lower (upper) boundary of the shaded bands for $v_2$ ($R_{\rm AA}$),
while $\frac{dN}{dg}$=1600 corresponds to upper (lower) boundary for $v_2$ ($R_{\rm AA}$).}
\end{figure}


Figure~\ref{fig:2} (a)-(b) shows the centrality dependence of $v_2$
in two high $\pT$ selections.  They are compared with four pQCD jet
quenching model calculations, AMY, HT and ASW from
\cite{Bass:2008rv} and WHDG from~\cite{Wicks:2005gt}.  The WHDG
model was calculated for gluon density $dN/dg=1000-1600$, a range
constrained by 0-5\% ($N_{\rm part}=351$) $\pi^0$ $R_{\rm AA}$
data~\cite{Adare:2008cg}.  The calculation assumes analytical
Woods-Saxon nuclear geometry with a longitudinal Bjorken expansion.
The AMY, HT and ASW models were fitted independently to the 0-5\%
$\pi^0$ $R_{\rm AA}$ data~\cite{Bass:2008rv}; they were implemented
in a 3D ideal hydrodynamic code with identical initial Wood-Saxon
nuclear geometry, medium evolution and fragmentation functions.  The
HT and ASW models include only coherent radiative energy loss,
while the AMY and WHDG also include collisional energy loss.  The
ASW and WHDG models predict quite sizable, and similar $v_2$, while
the HT and AMY models tend to give much smaller $v_2$.  However, all
models significantly under-predict the $v_2$ data in $6<\pT<9$
GeV/$c$ range.  For $\pT>9$ GeV/$c$, ASW and WHDG results show a
better agreement with the 20-30\% ($N_{\rm part}=167$) centrality
bin due to a slow decrease of $v_2$ with $\pT$ in this bin (see
Fig.~\ref{fig:1}(b)).  However, this seems to be accidental, since
the $v_2$ values for the other centrality bins remain large, and
are significantly above the WHDG calculations (the p-value for the
agreement is $<10^{-4}$).

In all these models, the inclusive suppression $R_{\rm AA}$ and
$v_2$ are anti-correlated, ${\it i.e.}$~a smaller $R_{\rm AA}$
implies a larger $v_2$ and vice versa.  Consequently, more
information can be obtained by comparing the data with a given
model for both $R_{\rm AA}$ and $v_2$.  Fig.~\ref{fig:2} (c)-(d)
compares the centrality dependence of $\pi^0$ $R_{\rm AA}$ data to
four model calculations for the same two $\pT$
ranges~\cite{Adare:2008qa}.  The calculations are available for a
broad centrality range for WHDG, but only in 0-5\% and 20-30\%
centrality bins for AMY, HT and ASW.  The level of agreement varies
among the models.  The HT calculations are slightly above the data
in the most central bin, while WHDG systematically under-predicts
the data over the full centrality range, though better agrement
with the data is obtained for $\pT > 9$ GeV/$c$.  On the other hand,
ASW and AMY calculations agree with the data very well in both
$\pT$ ranges.  The different levels of agreement among the models
are partially due to their different trends of $R_{\rm AA}$ with
$\pT$: WHDG and ASW results have stronger $\pT$ dependences than
what is impled by the data, and tend to deviate at low $\pT$ when
fitted to the full $\pT$ range~\cite{Adare:2008cg,Bass:2008rv}.
Given the larger fractional systematic error for $R_{\rm AA}$
measurements compared to the $v_2$ measurements, the deviation of
$v_2(N_{\rm part})$ from the data is more dramatic than that for
the $R_{\rm AA}(N_{\rm part})$.  Nevertheless, Fig.~\ref{fig:2}
clearly shows the importance for any model to simultaneously
describe the $R_{\rm AA}$ and the azimuthal anisotropy of the data.


The fact that the high $\pT$ $v_2$ at RHIC exceeds expectation of
pQCD jet-quenching models was first pointed out in
Ref.~\cite{Shuryak:2001me} in 2002.  This was not a serious issue
back then since the $\pT$ reach of early measurements was rather
limited, and the $v_2$ could be strongly influenced, up to 6
GeV/$c$ for pions, by collective flow and recombination effects
rather than jet quenching~\cite{Greco:2003mm}.  Fig.~\ref{fig:2}
clearly shows that the $v_2$ at $\pT$ above 6 or even 9 GeV/$c$
still exceeds the pQCD-based energy loss models.  It is possible
that geometrical effects due to fluctuations and CGC effects,
ignored in these models, can increase the calculated $v_2$; it is
also possible that the energy loss process in the sQGP has a
steeper $l$ dependence ({\it e.g.}~AdS/CFT) than what is currently
implemented in these models.

\begin{figure}[ht]
\includegraphics[width=1.0\linewidth]{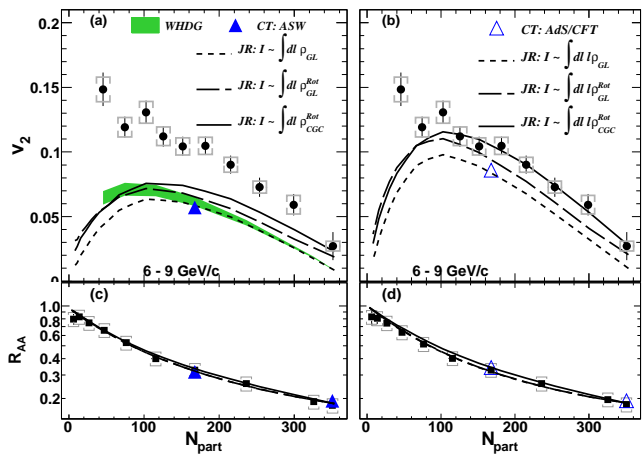}
\caption{\label{fig:3} (Color online) $v_2$ vs $N_{\rm part}$ in 6-9 GeV/$c$ compared with various models:
(a) WHDG~\cite{Wicks:2005gt} (shaded bands), ASW~\cite{Marquet:2009eq} (solid triangle), and three JR calculations~\cite{Jia:2010ee} with
quadratic $l$ dependence with longitudinal expansion for Glauber geometry (dashed lines),
rotated Glauber geometry(long dashed lines) and rotated CGC geometry (solid lines); (b) Same as (a)
except that AdS/CFT modified calculation in ASW framework (triangle) from~\cite{Marquet:2009eq} is shown and
the JR calculations were done for cubic $l$ dependence with longitudinal expansion;
(c)-(d) the comparison of calculated $R_{\rm AA}$s from these models with data.}
\end{figure}

To test whether these two ideas could bridge the difference between
data and theory, we compare the data with the JR model
from~\cite{Jia:2010ee}.  This model is based on a na\"ive jet
absorption picture with an exponential survival probability
$e^{-\kappa I}$ for jets, where the line integral $I=\int
dl\hspace{1mm}\rho$ is chosen for a quadratic dependence of
absorption in a longitudinally expanding medium, and $\kappa$ is
tuned to reproduce the central $R_{\rm AA}$ data.  The medium
density $\rho$ is given by two leading candidates of the initial
geometry: MC Glauber geometry $\rho_{\rm GL}(x,y)=0.43\rho_{\rm
part}(x,y)+0.14\rho_{\rm coll}(x,y)$, {\it i.e.}~a mixture of
participant density profile and binary collision profile from
PHOBOS~\cite{Back:2002uc}; and MC CGC geometry $\rho_{\rm
CGC}(x,y)$ of Dresher \& Nara~\cite{Drescher:2006pi}.  The effect of
fluctuations for both profiles were included via the standard
rotation procedure~\cite{Alver:2008zza}.  The short-dashed curves in
Fig.~\ref{fig:3}(a) show that the result for Glauber geometry
without rotation ($\rho_{\rm GL}$) compares reasonably well with
those from WHDG~\cite{Wicks:2005gt} and a version of ASW model
from~\cite{Marquet:2009eq}.  Consequently, we use the JR model to
estimate the shape distortions due to fluctuations and CGC effects.
The results for Glauber geometry with rotation ($\rho_{\rm GL}^{\rm
Rot}$) and CGC geometry with rotation ($\rho_{\rm CGC}^{\rm Rot}$)
each lead to an $\sim 15-20$\% increase of $v_2$ in mid-central
collisions.  However, these calculated results still fall below the
data.

Figure~\ref{fig:3}(b) compares the same data with three JR models
for the same matter profiles, but calculated for a line integral
motivated by AdS/CFT correspondence $I=\int dl \hspace{1mm}l\rho$.
The stronger $l$ dependence for $\rho_{\rm GL}$ significantly
increases (by $>50$\%) the calculated $v_2$, and brings it close to
the data for mid-central collisions.  However, a sizable fractional
difference in central bin seem to require additional increase from
fluctuations and CGC geometry.  Fig.~\ref{fig:3} (b) also shows a CT
model from~\cite{Marquet:2009eq}, which implements the AdS/CFT $l$
dependence within the ASW framework~\footnote{The two CT models
(ASW and AdS/CFT) in Fig.~\ref{fig:3} are based on a 3D ideal
hydrodynamic code slightly different from that of the ASW model
shown in Fig.~\ref{fig:2}.}; it compares reasonably well with the
JR model for $\rho_{\rm GL}$ (short-dashed curves).  Note that the
CT or JR models in Fig.~\ref{fig:3} have been tuned independently
to reproduce the 0-5\% $\pi^0$ $R_{\rm AA}$ data, and they all
describe the centrality dependence of $R_{\rm AA}$ very well (see
Fig.~\ref{fig:3} (c)-(d)).  On the other hand, these models predict
a stronger suppression for dihadrons than for single hadrons,
opposite to experimental findings~\cite{Adare:2010ry}, thus a
global confrontation of any model with all experimental observables
is warranted.

In summary, we presented results on $\pi^0$ azimuthal anisotropy
($v_2$) in $1<p_{T}<18$ GeV/$c$ in Au+Au collisions at
$\sqrt{s_{\rm NN}}$=200~GeV.  The measurements indicate sizable
$v_2(\pT)$ that decreases gradually for $3\lesssim \pT
\lesssim7-10$ GeV/$c$, but remains positive for $\pT>10$ GeV/$c$.
This large $v_2$ is striking in that it exceeds expectations of
pQCD energy-loss models even at $\pT\sim10$ GeV/$c$.  Estimates of
the $v_2$ increase due to modifications of initial geometry from
gluon saturation effects and fluctuations indicate that they are
insufficient to reconcile data and theory.  Incorporating an
AdS/CFT-like path-length dependence for jet quenching in a
pQCD-based framework~\cite{Marquet:2009eq} and a schematic
model~\cite{Jia:2010ee} both compare well with the data.  However,
more detailed study beyond these simplified models are required to
quantify the nature of the path-length dependence.  Our precision
data provide key experimental constraints on the role of initial
geometry and for elucidating the jet quenching mechanism.


We thank the staff of the Collider-Accelerator and
Physics Departments at BNL for their vital contributions.
We acknowledge support from
the Office of Nuclear Physics in DOE Office of Science and NSF (USA),
MEXT and JSPS (Japan),
CNPq and FAPESP (Brazil),
NSFC (China),
MSMT (Czech Republic),
IN2P3/CNRS and CEA (France),
BMBF, DAAD, and AvH (Germany),
OTKA (Hungary),
DAE and DST (India),
ISF (Israel),
NRF (Korea),
MES, RAS, and FAAE (Russia),
VR and KAW (Sweden),
U.S.  CRDF for the FSU,
US-Hungary Fulbright,
and US-Israel BSF.



\begin{thebibliography}{29} 

\bibitem{RHIC} K.~Adcox {\it et al.},  Nucl.\ Phys.\ A {\bf 757}
    (2005) 184;
 J.~Adams {\it et al.},
  Nucl.\ Phys.\ A {\bf 757} (2005) 102;
 B.~B.~Back {\it et al.},
 Nucl.\ Phys.\ A {\bf 757} (2005) 28;
I.~Arsene {\it et al.},
 Nucl.\ Phys.\ A {\bf 757} (2005) 1.

\bibitem{Gyulassy:2003mc}
  M.~Gyulassy, I.~Vitev, X.~N.~Wang and B.~W.~Zhang,
  arXiv:nucl-th/0302077.

\bibitem{Adcox:2001jp}
  K.~Adcox {\it et al.},
  Phys.\ Rev.\ Lett.\  {\bf 88}, 022301 (2002)

\bibitem{Adler:2002tq}
  C.~Adler {\it et al.}
  Phys.\ Rev.\ Lett.\  {\bf 90}, 082302 (2003)

\bibitem{Adare:2009vd}
  A.~Adare {\it et al.},
  Phys.\ Rev.\  C {\bf 80}, 024908 (2009)

\bibitem{Majumder:2010qh}
  A.~Majumder and M.~Van Leeuwen,
  arXiv:1002.2206

\bibitem{Adare:2008cg}
  A.~Adare {\it et al.},
  Phys.\ Rev.\  C {\bf 77}, 064907 (2008)

\bibitem{Bass:2008rv}
  S.~A.~Bass  {\it et al.}
  Phys.\ Rev.\  C {\bf 79}, 024901 (2009)

\bibitem{Muller:2008zzm}
  B.~Muller,
  Prog.\ Theor.\ Phys.\ Suppl.\  {\bf 174}, 103 (2008).

\bibitem{Peigne:2008wu}
  S.~Peigne and A.~V.~Smilga,
  Phys.\ Usp.\  {\bf 52}, 659 (2009)

\bibitem{Gubser:2008as}
  S.~S.~Gubser, D.~R.~Gulotta, S.~S.~Pufu and F.~D.~Rocha,
  JHEP {\bf 0810}, 052 (2008);

\bibitem{Miller:2007ri}
  M.~L.~Miller, K.~Reygers, S.~J.~Sanders and P.~Steinberg,
  Ann.\ Rev.\ Nucl.\ Part.\ Sci.\  {\bf 57}, 205 (2007)

\bibitem{Alver:2008zza}
  B.~Alver {\it et al.},
  Phys.\ Rev.\  C {\bf 77}, 014906 (2008)

\bibitem{Drescher:2006pi}
  H.~J.~Drescher and Y.~Nara,
  Phys.\ Rev.\  C {\bf 75}, 034905 (2007)

\bibitem{Luzum:2008cw}
  M.~Luzum and P.~Romatschke,
  Phys.\ Rev.\  C {\bf 78}, 034915 (2008)
  [Erratum-ibid.\ {\bf 79}, 039903 (2009)]

\bibitem{Hirano:2009ah}
  T.~Hirano and Y.~Nara,
  Phys.\ Rev.\  C {\bf 79}, 064904 (2009)

\bibitem{Adler:2005rg}
  S.~S.~Adler {\it et al.},
  Phys.\ Rev.\ Lett.\  {\bf 96}, 032302 (2006)

\bibitem{Adler:2006bw}
  S.~S.~Adler {\it et al.},
  Phys.\ Rev.\  C {\bf 76}, 034904 (2007)

\bibitem{Afanasiev:2009iv}
  S.~Afanasiev {\it et al.},
 Phys.\ Rev.\  C {\bf 80}, 054907 (2009)

\bibitem{Adare:2008cqb}
  A.~Adare {\it et al.},
  Phys.\ Rev.\  C {\bf 78}, 014901 (2008)

\bibitem{Adare:2008qa}
  A.~Adare {\it et al.},
  Phys.\ Rev.\ Lett.\  {\bf 101}, 232301 (2008)

\bibitem{Wicks:2005gt}
  S.~Wicks, W.~Horowitz, M.~Djordjevic and M.~Gyulassy,
  Nucl.\ Phys.\  A {\bf 784}, 426 (2007);

\bibitem{Shuryak:2001me}
  E.~V.~Shuryak,
  Phys.\ Rev.\  C {\bf 66}, 027902 (2002)

\bibitem{Jia:2010ee}
  A.~Drees, H.~Feng and J.~Jia,
  Phys.\ Rev.\  C {\bf 71}, 034909 (2005);
  J.~Jia and R.~Wei,
  arXiv:1005.0645

\bibitem{Back:2002uc}
  B.~B.~Back {\it et al.} 
  Phys.\ Rev.\  C {\bf 65}, 061901 (2002)

\bibitem{Marquet:2009eq}
  C.~Marquet and T.~Renk,
  Phys.\ Lett.\  B {\bf 685}, 270 (2010)

\bibitem{Greco:2003mm}
  V.~Greco, C.~M.~Ko and P.~Levai,
  Phys.\ Rev.\  C {\bf 68}, 034904 (2003);
  R.~J.~Fries, B.~Muller, C.~Nonaka and S.~A.~Bass,
  Phys.\ Rev.\  C {\bf 68}, 044902 (2003)

\bibitem{Adare:2010ry}
   A.~Adare~{\it et al.},
  arXiv:1002.1077 [nucl-ex].

\end{thebibliography}
\end{document}